\documentstyle[]{article}
\begin{document}
\oddsidemargin  -10pt
\evensidemargin -10pt
\topmargin -47pt
\headheight 12pt
\headsep 25pt
\footskip 75pt
\textheight 9in \textwidth 6.75in
\columnsep .375in
\columnseprule 0pt
\twocolumn \sloppy
\parindent 1em
\leftmargini 2em
\leftmargin
\leftmargini
\leftmarginv .5em
\leftmarginvi .5em
\flushbottom
\tolerance=1000
\begin{titlepage}
\begin{flushright}
CFP-IOP-98/13 \\ {\bf August 1998}
\end{flushright}

\vspace{3cm}

\begin{center}
{\Large \bf SU(N) SUSY GUTS WITH STRING REMNANTS:\\
MINIMAL SU(5) AND BEYOND}

\bigskip

\bigskip

{\large \bf Jon Chkareuli}

\smallskip

{\it Institute of Physics, Georgian Academy of Sciences \\
380077 Tbilisi, Georgia}
\end{center}

\bigskip

\bigskip

\begin{abstract}
A new superstring-motivated framework for a treatment of $SU(N)$ SUSY
GUTs is argued. We show that all the present difficulties of the
minimal supersymmetric $SU(5)$ model can successfully be overcome
with a new renormalizable reflection-invariant superpotential with
two-adjoint scalars, one of which is interpreted as a massive string
mode essentially decoupled from the low-energy particle spectra. This
superpotential is proved  to properly fix a mass ratio of the basic
adjoint scalar moduli, while its gauge-type reflection symmetry
essentially protects the model from gravitational smearing. The
significant heavy threshold effect related with the generic mass
splitting of adjoint moduli is shown to alter appropriately the
running of gauge couplings towards the realistic string-scale grand
unification. Furthermore, the extension of the superpotential to some
$SU(N)$ GUTs gives rise to, among many degenerate vacua, the missing
VEV vacuum configuration for the basic adjoint scalar, thus providing
a further clue to a doublet-triplet splitting problem, on the one
hand, and a family symmetry $SU(N-5)_F$ for quarks and leptons, on
the other. We predict the existence on a TeV scale two or three
families of pseudo-Goldstone bosons of type ($5+\bar{5}$) +
$SU(5)$-singlets depending on $SU(7)$ or $SU(8)$ GUT selected.
\end{abstract}

\vspace{3cm}

\begin{center} {\large \it A talk given at XXIX International
Conference on High Energy Physics, \\ Vancouver, B.C., Canada, July
23-29, 1998.} \end{center}

\end{titlepage}

\twocolumn

\section{Introduction}

Presently, leaving aside the non-supersymmetric Grand Unified
Theories which certainly contradict the experiment unless some
special extension of particle spectrum at intermediate scale(s) is
made \cite{1}, one can see that even SUSY GUTs seem to be far from
perfection. The problems, as they appear for the prototype
supersymmetric $SU(5)$ model (with a minimal content for matter,
Higgs and gauge bosons) \cite{2}, can conventionally be qualified as
phenomenological and conceptual.

The phenomenological ones include:\newline
{\bf (1)} The large value of the strong coupling $\alpha _{s}(M_{Z})$
predicted, $\alpha _{s}(M_{Z})>0.126$ for the effective SUSY scale $
M_{SUSY}<1$ TeV \cite{3} in contrast to the world average value \cite{4} $
\alpha _{s}(M_{Z})=0.119(2)$;\newline
{\bf (2)} The proton decay due to the color-triplet $H_{c}(\stackrel{\_}{H}
_{c})$ exchange \cite{2} at a rate largely excluded by a combination of the
detailed renormalization group (RG) analysis for the gauge couplings
\cite{5}
and the improved lower limits on the proton decay mode $p\to
\bar{\nu}K^{+}$
from Super-Kamiokande and on the superparticle masses from LEP2{\ \cite{6};
}
\newline
(3) An absence of the sizeable neutrino masses $m_{\nu }\geq 10^{-2}$ eV,
as
it can explicitly be derived from the atmospheric neutrino deficit data
reported{\ \cite{7}, at least for one of neutrino species.}

Furthermore, for conceptual reasons, the present status of the minimal $
SU(5) $ appears to be inadequate as well:\newline
{\bf (4)} The first one is, of course, the doublet-triplet splitting
problem
entirely underlying the gauge hierarchy phenomenon in SUSY GUTs \cite{2};
\newline
{\bf (5)} The next point is an absence in the minimal theory of any flavor
or family symmetry mechanism which can guarantee together with rather
peculiar masses of quarks and leptons a nearly uniform mass spectrum for
their superpartners with a high degree of flavor conservation in SUSY
theories \cite{8};\newline
{\bf (6)} Then a low unification scale $M_{U}$ whose value lies one order
of
magnitude below than the typical string scale $M_{STR}\simeq 5\cdot
10^{17}$
GeV \cite{9};\newline
{\bf (7)} And lastly, the gravitational smearing of its principal
predictions (particularly for $\alpha _{s}(M_{Z})$) due to the
uncontrollable high-dimension operators induced by gravity in the
kinetic terms of the basic gauge SM bosons \cite{10} that makes the
ordinary $SU(5)$ model to be largely untestable.

The question arises: Where the possible solution to those problems could
come
from? According to a superstring theory \cite{9} we seem to say "no"
for turning into play of any new states other than the massive string
modes, or any mass scales besides the string one. The only exception
could be made for the adjoint scalar $\Sigma $ moduli states $\Sigma
_{8}(8_{c},1)$ and $\Sigma _{3}(1,3_{w})$ (in a self-evident
$SU(3)_{C}\otimes SU(2)_{W}$ notation) appearing in many string
models at a well-motivated intermediate scale $ m\sim
M_{P}^{2/3}M_{SUSY}^{1/3}$ \cite{11}. That means, as an example, the
starting mass of the basic adjoint scalar supermultiplet $\Sigma (24)$ of $
SU(5)$ theory (remaining just the non-Goldstone remnants $\Sigma _{8}$ and
$
\Sigma _{3}$ after symmetry breaking) may be taken at a scale $m$. Thus,
the
problems listed above remain to be addressed to some new superpotential
provided there is the "better" one that develops a proper vacuum
configurations
when including extra massive ($\sim M_{P}$) states and/or increasing its
symmetry. Strange as it may seem, those simple string-motivated rules of an
"allowed" generalization of the minimal $SU(5)$ are turned out to
work.

Towards this end, let us consider a general $SU(N)$ invariant
renormalizable
superpotential of two adjoint scalars $\Sigma $ and $\Omega $ satisfying
also the gauge-type reflection symmetry ($\Sigma \to -\Sigma $, $\Omega \to
\Omega $) inherited from superstrings
\begin{equation}
W=\frac{1}{2}m\Sigma ^{2}+\frac{1}{2}M_{P}\Omega ^{2}+\frac{1}{2}h\Sigma
^{2}\Omega +\frac{1}{3}\lambda \Omega ^{3}+W_H  \label{1}
\end{equation}
where the second adjoint $\Omega $ can be considered as a state originated
from the massive string mode with the (conventionally reduced) Planck mass
$
M_{P}=(8\pi G_{N})^{-1/2}\simeq 2.4\cdot 10^{18}$ GeV, while the basic
adjoint $\Sigma$ remains (relatively) light when one goes from the string
scale to
lower energies, $m\sim M_{P}^{2/3}M_{SUSY}^{1/3}$ \cite{11}. The
superpotential also includes the ordinary Higgs-doublet containing
fundamental chiral supermultiplets $H$ and $\bar{H}$ presented in
$W_H$ which is unessential for the moment.

We show below that the superpotential (\ref{1}) entirely constructed
according to our simple rules lead to the natural string scale unification
even in a case of the minimal $SU(5)$, whereas its new missing VEV vacuum
configurations in the higher $SU(N)$ symmetry cases can give a
further clue to other problems mentioned above: doublet-triplet
splitting, family symmetry, neutrino masses etc. At the same time,
due to the gauge reflection symmetry of the superpotential $W$, the
operators gravitationally induced at a Planck scale for the basic
adjoint $\Sigma $ (developing the principal VEV in the model, see
below) in the kinetic terms of the SM gauge bosons should have
dimension 6 and higher. Thus, their influence on our predictions
seems to be negligible in contrast to the standard $SU(5)$ where they
can largely be smeared out \cite{10}.

\section{The SU(5) model}

We start by recalling that, to one-loop order, gauge coupling
unification is given by the three RG equations relating the values of
the gauge couplings at the Z-peak $\alpha _{i}(M_{Z})$ $(i=1,2,3)$,
and the common gauge coupling $\alpha _{U}$ \cite{1}:
\begin{equation}
\alpha _{i}^{-1}=\alpha _{U}^{-1}+\sum_{p}\frac{b_{i}^{p}}{2\pi }ln\frac{
M_{U}}{M_{p}}  \label{2}
\end{equation}
where $b_{i}^{p}$ are the three b-factors corresponding to the $SU(5)$
subgroups $U(1)$, $SU(2)$ and $SU(3)$, respectively, for the particle
labeled by $p$. The sum extends over all the contributing particles in the
model, and $M_{p}$ is the mass threshold at which each decouples. All of
the
SM particles and also the second Higgs doublet of MSSM are already
presented
at the starting scale $M_{Z}$. The next is assumed to be
supersymmetric threshold associated with the decoupling of the
supersymmetric particles at some single effective scale
$M_{SUSY}$ \cite{3}; we propose thereafter the relatively low values
of $M_{SUSY}$, $M_{SUSY}\sim M_{Z}$ to keep sparticle masses
typically in a few hundred GeV region. The superheavy states, such as
the adjoint fragments $\Sigma _{8}$ and $\Sigma _{3}$ at the masses
$M_{8}$ and $M_{3}$, respectively, and the color-triplets $H_{c}$ and
$\bar{H_{c}}$ at a mass $M_{c}$ are also included in the evolution
equations (2). As to the superheavy gauge bosons and their superpartners
(X-states), they do not contribute to the Eq.(\ref{2}), for they are
assumed
to lie on the GUT scale $M_{U}$ ($M_{X}=M_{U}$), above which all particles
fill complete $SU(5)$ multiplets.

Now, by taking the special combination of Eqs.(\ref{2}) we are led to the
simple relation between gauge couplings and the logarithms of the
neighboring threshold mass ratios
\begin{eqnarray}
12\alpha _{2}^{-1}-7\alpha _{3}^{-1}-5\alpha _{1}^{-1}~=  \nonumber \\
\frac{3}{2\pi
}(-2ln\frac{M_{X}}{M_{c}}+ln\frac{M_{c}}{M_{3}}-7ln\frac{M_{3}
}{M_{8}}-\frac{19}{6}ln\frac{M_{SUSY}}{M_{Z}})
\label{3}
\end{eqnarray}
which can be viewed as the basis for giving the qualitative constraints to
the $\alpha _{s}(M_{Z})$ depending on the present (very precise)
measurement
of $sin^{2}\theta _{W}$ \cite{4} and superheavy mass splitting, when one
goes beyond the MSSM limit ($M_{X}=M_{c}=M_{3}=M_{8}$). One can see from
Eq.(
\ref{3}) that $\alpha _{s}$ increases with $\frac{M_{c}}{M_{3}}$ and
decreases with $\frac{M_{X}}{M_{c}}$, $\frac{M_{SUSY}}{M_{Z}}$ and,
especially, with $\frac{M_{3}}{M_{8}}$ (the largest coefficient before
logarithm). Unfortunately, in the standard $SU(5)$ case \cite{2} with
the degenerate adjoint moduli $\Sigma _{3}$ and $\Sigma _{8}$
($M_{3}=M_{8}$ at the GUT scale $M_{X}$) one is inevitably lead to
unacceptably high values of $\alpha _{s}(M_{Z})$ for the allowed
$M_{c}$ region \cite{5,6} and subTeV $ M_{SUSY}$ area \cite{3}.

However, a drastically different unification picture appears when a
generically large mass splitting between $\Sigma _{3}$ and $\Sigma _{8}$
that follows from a new superpotential (\ref{1}) is taken into account.
Actually, one can see that in a basic vacuum which breaks $SU(5)$ to
$ SU(3)_{C}\otimes SU(2)_{W}\otimes U(1)_{Y}$ the VEVs of the
adjoints $\Sigma $ and $\Omega $ are given by \begin{equation} \Sigma
(\Omega )=\frac{\sqrt{8mM_{P}}}{h}(\frac{2m}{h})
diag(1,~1,~1,~-3/2,~-3/2)
\label{4}
\end{equation}
respectively, with the hierarchically large VEV ratio being inverse to
their
masses, $\Sigma /\Omega =(2M_{P}/m)^{1/2}$. As this takes place, the
(physical) mass ratio of the survived adjoint moduli $\Sigma _{3}$ and $
\Sigma _{8}$ is turned out to be fixed (at a GUT scale)
\begin{equation}
M_{3}=10m~,~~~M_{8}=\frac{5}{2}m~,~~~\frac{M_{3}}{M_{8}}=4  \label{5}
\end{equation}
(just as in a non-supersymmetric $SU(5)$ case) in contrast to
$M_{3}/M_{8}=1$
in the standard one-adjoint superpotential \cite{2}.

So, with the observations made we are ready now to carry out the standard
two-loop analysis (with conversion from $\overline{MS}$ scheme to
$\overline{
DR}$ one included) \cite{1,12} for gauge ($\alpha _{1}$, $\alpha _{2}$, $
\alpha _{3}$) and Yukawa ($\alpha _{t}$, $\alpha _{b}$ and $\alpha _{\tau
}$
in a self-evident notation for top- and bottom-quarks and tau-lepton)
coupling evolution depending on, apart from the single-scale ($M_{SUSY}$)
supersymmetric threshold corrections mentioned above, the heavy $\Sigma $
moduli threshold only. This varies, in turn, from the GUT scale $M_{X}$ ($
M_{3}=M_{X}$, $M_{8}=\frac{1}{4}M_{X}$) down to some well-motivated
intermediate value $O(10^{14})$ GeV \cite{11} pushing thereafter the
$M_{X}$
up to the string scale $M_{STR}$. The mass splitting between weak triplet $
\Sigma _{3}$ and color octet $\Sigma _{8}$ in themselves noticeably
decreases, while $M_{3}$ and $M_{8}$ run from $M_{X}$ down to the lower
energies, as it results from their own two-loop RG evolution, which is also
included in the analysis. On the other hand, the color triplets
$H_{c}(\bar{H
}_{c})$ are always taken at $M_{X}$, for the strings seem to say nothing
why
any extra states, other than the adjoint moduli $\Sigma _{3}$ and $\Sigma
_{8}$, could left relatively light.

Our results, as appeared after numerical integration of all the RG
equations
listed above, are largely summarized in Figure 1. One can see that
the $\alpha _{s}(M_{Z})$ values predicted (with a percent accuracy due to
the very precise value of $\sin ^{2}\theta _{W}$ $=0.2313(3)$ \cite{4} and
top-Yukawa coupling appropriately fixed at $M_{X}$), are in a good
agreement
with the world average value (see above) in contrast to the standard SUSY $
SU(5)$ taken under the same conditions.

Remarkably enough, the presently testable (SUSY threshold neglecting)
top-bottom unification \cite{13} turned out to work well in the model, thus
giving the good prediction of the top-quark mass. Furthermore, the low
starting values of $\alpha _{t}$ and $\alpha _{b}$ at $M_{X}$ in this case,
as well as the closeness of the unification mass $M_{X}$ to the string
scale, allow one to make a next step towards the most symmetrical case
which
can be realized in the present string-motivated $SU(5)$ - Yukawa and gauge
coupling superunification at a string scale:
\begin{eqnarray}
\alpha _{t}(M_{X}) &=&\alpha _{b}(M_{X})=\alpha _{\tau }(M_{X})=\alpha
_{U}~,
\nonumber \\
M_{X} &=&M_{STR}  \label{6}
\end{eqnarray}
This conjecture certainly could concern the third-family Yukawa couplings
solely, since those ones could naturally arise from the basic
string-inspired interactions, whereas masses and mixing of the other
families seemed to be caused by some more complex and model-dependent
dynamics showing itself at lower energies. Due to a crucial reduction of a
number of the fundamental parameters the gauge-Yukawa coupling unification
leads immediately to a series of the very distinctive predictions (of $
\alpha _{s}$, in general, and masses in absence of any large supersymmetric
threshold corrections)
\begin{eqnarray}
\alpha _{s}(M_{Z}) &=&0.119\pm 0.001~,~~m_{t}=180\pm 1~,  \nonumber \\
\frac{m_{b}}{m_{\tau }} &=&1.79\pm 0.01~,~~tan\beta =52\pm 0.2  \label{7}
\end{eqnarray}
in a surprising agreement with experiment \cite{4}. In Figure 2 the
superunification of gauge and Yukawa couplings is demonstrated.

This is how a new superpotential (\ref{1}) seems for the first time to open
the way to the natural string-scale grand unification in the supersymmetric
$
SU(5)$, as prescribed at low energies by the gauge coupling values and the
minimal particle content \cite{14}.

\section{Beyond the SU(5)}

As a general analysis of the superpotential $W$ (\ref{1}) shows \cite{15}
that possible VEV patterns of the adjoints $\Sigma $ and $\Omega $
include the following four cases only: (i) the trivial symmetry
unbroken case, $\Sigma = \Omega = 0$; (ii) the single-adjoint
condensation, $\Sigma =0$, $\Omega \neq 0$; (iii) the "parallel"
vacuum configurations, $\Sigma \propto\Omega$ and (iv) the
"orthogonal" vacuum configurations, $Tr(\Sigma \Omega )=0$. While the
Planck-mass mode $\Omega$ having a cubic term in $W$ develops in all
non-trivial cases only a "standard" VEV pattern which breaks the
starting $SU(N)$ symmetry to $SU(k)\otimes S(N-k)\otimes U(I)$ the
basic adjoint $\Sigma$ develops a radically new missing VEV vacuum
configuration in a case (iv), thus giving a "double" breaking of
$SU(N)$ to $SU(k/2)\otimes SU(k/2)\otimes SU(N-k)\otimes
U(I)_1\otimes U(I)_2$ in this case:  \begin{eqnarray}
~k~~~~~~~~~\hspace{0.5cm}
N-k~~~~~~~\hspace{0.5cm}
\nonumber \\
\Omega =\omega diag(\overbrace{~1~...~1~}~,
~\overbrace{-\frac{k}{N-k}...-\frac{k}{N-k}})
\nonumber \\
~~~~~k/2~~~~\hspace{0.5cm}
k/2~~~~~~~\hspace{0.5cm}N-k
\nonumber \\
\Sigma =\sigma diag(\overbrace{~1~...~1~}~,
\overbrace{~-1~...~-1~}~, \overbrace{~0~...~0~})
\label{8}
\end{eqnarray}
with $\omega =(-m/h)(\frac{N-k}{N-2k})$ and $\sigma = \sqrt{mM_P}/h
(\frac{2N}{N-k})^{1/2}$, respectively.

Now, if it is granted that the "missing VEV subgroup" $SU(N-k)$ is just
the weak symmetry group $SU(2)_W$, we come to a conclusion that the numbers
of fundamental colors and flavors (or families) must be equal
($n_C=n_F=k/2$). They all are entirely unified in the framework of
the starting $SU(8)$ symmetry \cite{15}. The quark-lepton families
(and their superpartners) having been properly assigned to $SU(8)$
multiplets \cite{15} appear as the fundamental triplet of the chiral
family symmetry $SU(3)_F$ \cite{16} that meets a natural
conservation of flavor both in the particle and sparticle sectors,
respectively \cite{8}.  Another possibility is when $SU(N-k)$
identifying with the color symmetry group $SU(3)_C$, thus giving
interrelation between the weak and flavor groups ($n_W=n_F=k/2$) in
the framework of the starting $SU(7)$ symmetry. In the latter case
the quark-lepton (squark-slepton) families are doublet plus singlet
under the family symmetry $SU(2)_F$ providing the valuable mass
matrices for quarks and leptons \cite{15} as well as a flavor
conservation in the sparticle sector \cite{8}. The higher $SU(N)$
groups, if considered, are based solely on those principal
possibilities mentioned.

Let us see now how this missing VEV mechanism works to solve
doublet-triplet splitting problem in $SU(8)$ or $SU(7)$ GUT due to
the superpotential $W$ (\ref{1}). There the $W_H$ part is, in fact,
the only reflection-invariant coupling of the basic adjoint $\Sigma$
with a pair of the ordinary Higgs-boson containing supermultiplets
$H$ and $\bar{H}$ \begin{equation} W_H=f\bar{H}\Sigma H~~~~(\Sigma
\to -\Sigma,~\bar{H}H\to -\bar{H}{H}) \label{9} \end{equation} having
the zero VEVs, $\bar{H}=H=0$, during the first stage of the symmetry
breaking. Thereupon $W_H$ turns to the mass term of $H$ and $\bar{H}$
depending on the missing VEV pattern (\ref{8}). This vacuum, while
giving generally heavy masses (of the order of $M_{GUT}$) to them,
leaves their weak components strictly massless. To be certain we
must specify the multiplet structure of $H$ and $\bar{H}$ in the both
cases of the weak-component and color-component missing VEV vacuum
configurations, that is, in $SU(8)$ and $SU(7)$ GUTs, respectively.
In the $SU(8)$ case $H$ and $\bar{H}$ are fundamental octets whose
weak components (ordinary Higgs doublets) do not get masses from the
basic coupling (\ref{9}). In the SU(7) case $H$ and $\bar{H}$ are the
2-index antisymmetric 21-plets which (after the proper projecting out
of extra states) contain just a pair of the massless Higgs doublets.
Thus, there certainly is a natural doublet-triplet splitting in the
both cases although we drive at the vanishing $\mu$ term on this
stage. However, one can argue that the right order $\mu$ term always
appears from the radiative corrections on the next stage when SUSY
breaks \cite{15}.

Inasmuch as the missing VEV vacua appear only in the higher than
$SU(5)$ symmetry cases and extra flavor symmetry should break
\begin{equation}
SU(N-5)_F\otimes U(I)_1\otimes U(I)_2 \to U(1)_Y
\label{10}
\end{equation}
at the GUT scale as well (not to spoil the gauge coupling unification),
a question arises: How can those vacua survive so as to be subjected at
most
to the weak scale order shift? This requires, in general, that a
superpotential (\ref{1}) to be strictly protected from any large
influence of $N-5$ scalars $\varphi ^{(n)}$ ($n=1,...,N-5$) providing
the flavor symmetry breaking (\ref{10}). Technically, such a
custodial symmetry could be the superstring-inherited anomalous
$U(1)_A$ \cite{17} which can naturally get untie those two sectors
and induce the right-scale flavor symmetry breaking (\ref{10})
through the Fayet-Iliopoulos D-term \cite{2}. Anyway, as it takes
place in the supersymmetric $SU(N)$ theory considered, the
accidental, while radiatively broken, global symmetry $SU(N)_I\otimes
SU(N)_{II}$ appears and $N-5$ families of the pseudo-Goldstone states
of type \begin{equation} 5~+~\bar{5}~+~SU(5)-{\rm singlets}
\label{11} \end{equation} are produced at a TeV scale where SUSY
softly breaks.  Together with ordinary quarks and leptons and their
superpartners the two or three families of PG states (\ref{11}),
depending on $SU(7)$ or $SU(8)$ GUT selected, determine the whole
particle environment at low energies.

\section{Conclusion}

The recent Super-Kamiokande data \cite{7} arise a question about a
modification of the SM to get a mass of order 0.1 eV at least for one
of neutrino species. That means, in general, a particle
content of the SM or the minimal $SU(5)$ should be extended to
include new states, that is, the properly heavy right-handed
neutrinos or even light sterile left-handed ones. We have found that
the proper missing VEV vacuum configurations require the extended
GUTs $SU(7)$ or $SU(8)$ where such states, together with ordinary
quarks and leptons, naturally appear, thus providing on this stage at
least a qualitative explanation \cite{15} for data \cite{7}.

Meanwhile, despite the common origin there is a principal
difference between the $SU(7)$
and $SU(8)$ cases that manifests itself not only in number of PG
families ({\ref{11}). The point is the basic adjoint $\Sigma$ moduli
mass ratio $M_3/M_8$ (which, as we could see in the $SU(5)$,
essentially determines a high-energy behavior of gauge couplings)
appears according to the missing VEV vacua (\ref{8}) to be 2 and 1/2
for $SU(7)$ and $SU(8)$, respectively. That means the unification
scale in $SU(7)$ can be pushed again to the string scale, while scale in
$SU(8)$ always ranges closely to the standard unification value \cite{15}.
The detailed RG analysis \cite{15} carried along with that of $SU(5)$
(Section 2) lives up to our expectations.  So, the $SU(7)$ seems to be
the only GUT that can give a solution to all seven problems
which we have started from.

\section*{Acknowledgments}

I would like to acknowledge the stimulating conversations with many
colleagues, especially with Alexei Anselm, Riccardo Barbieri,
Zurab Berezhiani, Gia Dvali, Colin Froggatt, Mike Green, S.
Randjbar-Daemi, Alexei Smirnov, David Sutherland, and not least, with
my collaborators Ilia Gogoladze and Archil Kobakhidze. Financial
support by INTAS Grants No. RFBR 95-567 and 96-155 are also
gratefully acknowledged.

\newpage

\section*{Figure captions}

\bigskip

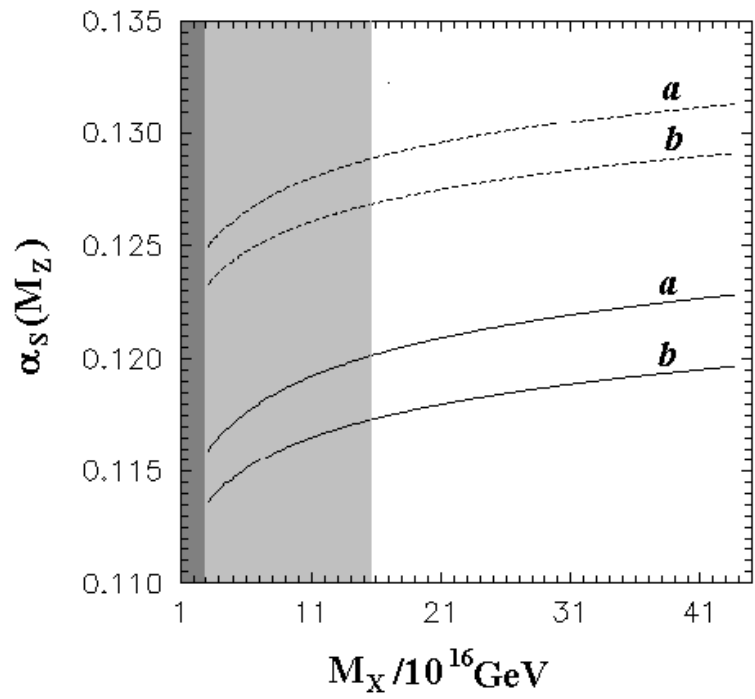
{\bf Figure 1:}~ The predictions in the present
model (the solid lines) and in the standard supersymmetric $SU(5)$ model
(the dotted lines) of $\alpha _{s}(M_{Z})$ as a function of the grand
unification scale $M_{X}$ for the two cases of small $tan\beta $ values (a)
with top-Yukawa coupling $\alpha _{t}(M_{X})=0.3$ taken and large $tan\beta
$
(b) values corresponding to top-bottom unification under $\alpha
_{t}(M_{X})=\alpha _{b}(M_{X})=0.05$. The unification mass $M_{X}$ varies
from the adjoint-moduli threshold-degeneration point ($M_{X}=M_{\Sigma }$)
to the string scale ($M_{X}=M_{STR}$ with a level $k=1$), while the
color-triplet mass is assumed to be at unification scale in all cases ($
M_{c}=M_{X}$). The all-shaded areas on the left are generally disallowed by
the present bound \protect\cite{4} on nucleon stability for both cases
((a),
dark) and ((b), light), respectively.

\medskip

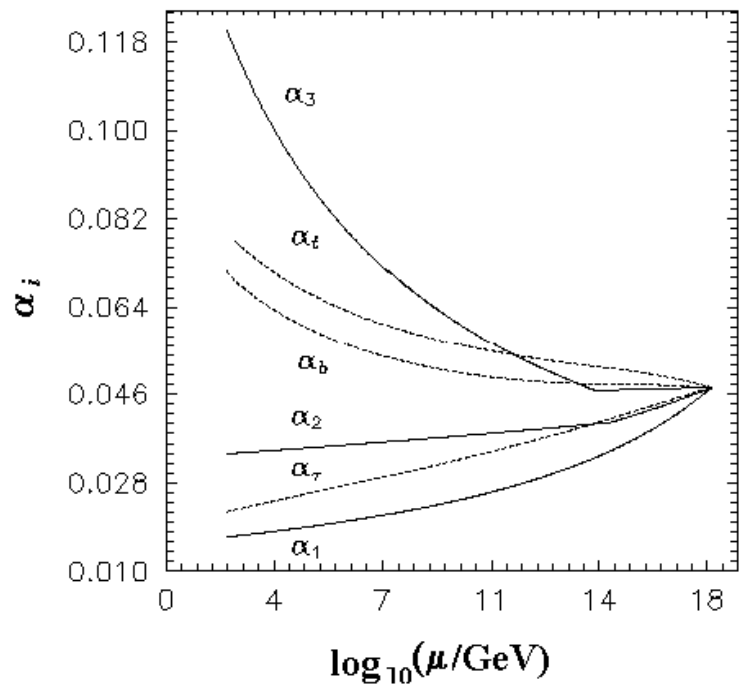
\noindent{\bf Figure 2:}~ The superunification of gauge ($\alpha
_{1}$, $\alpha _{2}$, $\alpha _{3}$) and Yukawa ($\alpha _{t}$,
$\alpha _{b}$, $ \alpha _{\tau }$) couplings at the string scale (the
solid and dotted lines, respectively).

\begin{thebibliography}{99}
\bibitem{1}  W. de Boer, {\em Prog. Part. Nucl. Phys.} 33, 201 (1994).

\bibitem{2}  S. Dimopoulos and H. Georgi, {\em Nucl. Phys.}
B 193, 150 (1981); N. Sakai, {\em Z. Phys.}
C 11, 153 (1981); S. Weinberg, {\em Phys. Rev.} D 26, 287 (1982);
N. Sakai and T. Yanagida, {\em Nucl. Phys.} B 197, 533 (1982).

\bibitem{3}  J. Bagger, K. Matchev and D. Pierce, {\em Phys. Lett.}
B 348, 443 (1995);
P. Langacker and N. Polonsky, {\em Phys. Rev.} D 52, 3081 (1995).

\bibitem{4}  Particle Data Group, {\em Eur. Phys. J. C.} 3, 1 (1998).

\bibitem{5}  H. Hisano, T. Moroi, K. Tobe and T. Yanagida,
{\em Mod. Phys. Lett.} A 10, 2267 (1995).

\bibitem{6}  H. Murayama, {\it Nucleon Decay in GUT and Non-GUT SUSY
models}
, LBNL-39484; hep-ph/9610419.

\bibitem{7}The Super-Kamiokande Coll. (Y.Fukuda {\it et al}.),
{\em Phys. Rev. Lett.} 81, 1562 (1998).

\bibitem{8}  M. Dine, R. Leigh and A. Kagan,
{\em Phys. Rev.} D 48, 4269 (1993);
L.J. Hall and H. Murayma,
{\em Phys. Rev. Lett.} 75, 3985 (1995); Z. Berezhiani,
{\em Phys. Lett.} B 417, 287 (1998).

\bibitem{9}  For a recent discussion and extensive references, see K.
Dienes, {\em Phys. Rep.} 287, 447 (1997).

\bibitem{10}  C.T. Hill, {\em Phys.Lett.} 135 B, 47 (1984);
L.J. Hall and U. Sarid, {\em Phys. Rev. Lett.} 70, 2673 (1993);
S. Urano, D. Ring and R. Arnowitt, {\em Phys. Rev. Lett.} 76, 3663 (1996).

\bibitem{11}  A.E. Faraggi, B. Grinstein and S. Meshkov, {\em Phys. Rev.}
D 47, 5018 (1993); C. Bachas, C. Fabre and T. Yanagida,
{\em Phys. Lett.} B 370, 49 (1996);
M. Bastero-Gil and B. Brahmachari, {\em Phys. Lett.} B403, 51 (1997).

\bibitem{12}  V. Barger, M.S. Berger and P. Ohmann,
{\em Phys. Rev.} D 47, 1093 (1993);
P. Martin and M.T. Vaughn, {\em Phys. Rev.} D 50, 2282 (1994).

\bibitem{13}  L.J. Hall, R. Rattazzi and U. Sarid,
{\em Phys. Rev.} D 50, 7048 (1994);
M. Carena, M. Olechowski, S. Pokorski and C.E.M. Wagner, {\em Nucl. Phys.}
B 426, 269 (1994).

\bibitem{14}  J.L. Chkareuli and I.G. Gogoladze, {\em Phys. Rev.} D 58, 551
(1998).

\bibitem{15}  J.L. Chkareuli and A.B. Kobakhidze, {\em Phys. Lett.}
B 407, 234 (1997); J.L. Chkareuli, I.G. Gogoladze and
A.B. Kobakhidze, {\em Phys. Rev. Lett.} 80, 912 (1998); {\em Phys.
Lett.} B (in press).

\bibitem{16}  J.L. Chkareuli, {\em JETP Letters} 32, 671 (1980); Z.G.
Berezhiani, {\em Phys. Lett.} B 129, 99 (1983); B 150, 117 (1985);
J.L. Chkareuli, {\em Phys. Lett.} B 246, 498 (1990); B 300, 361
(1993).

\bibitem{17}  M. Green and J. Schwarz, {\em Phys. Lett.} B 149, 117
(1998).
\end{thebibliography}
\end{document}